\documentstyle[12pt,aps,epsfig,psfig]{revtex}
\newcommand{\be}{\begin{equation}}
\newcommand{\ee}{\end{equation}}
\newcommand{\bea}{\begin{eqnarray}}
\newcommand{\eea}{\end{eqnarray}}

\begin{document}
\vspace*{2cm}
\begin{center}
{\Large\bf The directed flow maximum near $c_s=0$}
\\[2cm]
{\bf J.\ Brachmann$^1$, A.\ Dumitru$^2$, 
H.\ St\"ocker$^1$, W.\ Greiner$^1$}
\\[0.4cm]
{\small $^1$Institut f\"ur Theoretische Physik der 
J.W.\ Goethe-Universit\"at}\\
{\small Postfach 111932, D-60054 Frankfurt am Main, Germany}
\\[0.2cm]
{\small $^2$Physics Department, Columbia University}\\
{\small 704 Pupin Hall, 538W 120th Street, New York, NY 10027, USA}
\\[1cm]
\end{center}
\vspace*{1.5cm}
\begin{abstract}
We investigate the excitation function of quark-gluon plasma formation
and of directed in-plane flow of nucleons in the energy range
of the BNL-AGS and for the $E^{kin}_{Lab}=40A$~GeV Pb+Pb collisions
performed recently
at the CERN-SPS. We employ the three-fluid model with dynamical
unification of kinetically equilibrated fluid elements.
Within our model with first-order phase transition at high
density, droplets of QGP coexisting with hadronic matter
are produced already at BNL-AGS energies, $E^{kin}_{Lab}\simeq 10A$~GeV.
A substantial decrease of the isentropic velocity of sound, however,
requires higher energies, $E^{kin}_{Lab}\simeq40A$~GeV.
We show the effect on the flow of nucleons in the reaction plane.
According to our model calculations, kinematic
requirements and EoS effects work hand-in-hand at $E^{kin}_{Lab}=40A$~GeV
to allow the observation of the dropping velocity of sound
via an {\em increase} of the directed flow
around midrapidity as compared to top BNL-AGS energy.
\end{abstract}
\newpage

The theory of strong interactions, QCD, exhibits a
thermodynamical phase transition to a so-called quark-gluon plasma (QGP)
at high energy density~\cite{QCDpt}. To achieve densities far beyond that
of the nuclear ground state one collides heavy ions at high
energies~\cite{shockaus,oldflow,Clare:1986qj,LBCs,Yaris,Hung}.
If the equation of state (EoS) exhibits no anomalous structure,
one expects that the volume and life-time of the QGP increase with energy.

Due to non-equilibrium effects, however,
higher impact energies do not necessarily yield a larger amount of
energy deposition in the central region. One-fluid dynamical calculations
of the compression neglect initial non-equilibrium processes and
predict sharp signals for the phase transition to the
QGP~\cite{shockaus,oldflow,Clare:1986qj,LBCs,Yaris,Hung}.
How strongly these predictions are washed out by non-equilibrium
effects has to be investigated via excitation functions.
Therefore, the energy range from BNL-AGS energies ($2-11A$~GeV) to CERN-SPS
energies (previously only $160-200A$~GeV) is highly interesting to
study the EoS of nuclear matter, and possibly the onset of QGP formation.
In particular, data at $40A$~GeV has been taken recently at the CERN-SPS
to complete the already existing data~\cite{AGSdata,Kampert,Harry}
on the excitation function of collective flow.

Motivated by the fact that the proton rapidity distribution in high energy
$pp$
reactions is strongly forward-backward peaked~\cite{pp24}, we developed a
fluid-dynamical model in which the projectile and
target nucleons constitute {\em distinct} fluids. The continuity equations
for the energy-momentum tensor and net baryon current, $\partial_\mu
T_i^{\mu\nu}=S_i^\nu$, $\partial_\mu N_i^{\mu}=S_i$, are solved for each
of the fluids $i$. The couplings $S_i^\nu$, $S_i$ between those
fluids are obtained from a parametrization of
binary $NN$ collisions~\cite{2fluid}, which leads to a gradual deceleration
instead of instantaneous stopping as in the one-fluid model.
In the three-fluid model~\cite{paper1,PRC95} we
also consider the newly produced particles around midrapidity as a
distinct fluid, which we call ``fireball'', as
those particles populate a distinct rapidity region as well (at early times~!).
The assumption of local equilibrium is imposed
within each fluid, but the total energy-momentum tensor $T^{\mu\nu}=
T_1^{\mu\nu}+T_2^{\mu\nu}+T_3^{\mu\nu}$ does {\em not}
need to be that of an ideal fluid.

However, after several collisions projectile and target 
nucleons are stopped, and
local thermal equilibrium establishes between projectile and 
fireball fluid or between target and fireball fluid, respectively,
and finally for all three fluids.
The respective fluids $i$ and $j$ are then unified by 
adding their energy-momentum tensors and net-baryon currents at the 
corresponding space-time point,
$T_i^{\mu \nu}(x) + T_j^{\mu \nu}(x) = T^{\mu \nu}_{\rm unified}(x)$,
$N^\mu_i(x) + N^\mu_j(x) = N^\mu_{\rm unified}(x)$.
Common values for $e$, $p$, $\rho$ and $u^\mu\equiv\gamma(1,{\bf v})$
are then obtained from
$T^{\mu \nu}_{\rm unified} = (e+p)\, u^\mu u^\nu - p \, g^{\mu\nu}$ and
$N^\mu_{\rm unified}  = \rho\, u^\mu$, and the given EoS $p=p(e,\rho)$.
The dynamical unification allows friction-free
ideal expansion of equilibrated fluids.

The initial condition corresponds to two approaching but still separated
nuclei in the ground state. The model calculation
treats the compression (deceleration) stage
as well as the subsequent expansion. The fluid-dynamical equations are solved
in 3+1 dimensions, cf.~\cite{paper1} for
details. The EoS employed here exhibits a first order phase transition to a
QGP~\cite{Yaris}. The hadronic phase consists of nucleons
interacting via relativistic mean-fields~\cite{GorEOS},
plus thermal pions. The QGP phase is described within
the framework of the MIT-Bag model~\cite{MIT}
as an ideal gas of $u$ and $d$ quarks
and gluons, with a bag para\-meter $B^{1/4}=235$~MeV, resulting in a critical
temperature $T_c\simeq 170$~MeV at $\rho=0$, while the critical baryon-chemical
potential is $\mu_c=1.8$~GeV at $T=0$. The first
order phase transition is constructed via
Gibbs' conditions of phase coexistence.

\begin{figure}[p]
\vspace*{-1cm}
\hspace{0cm}\centerline{\hbox
{\psfig{figure=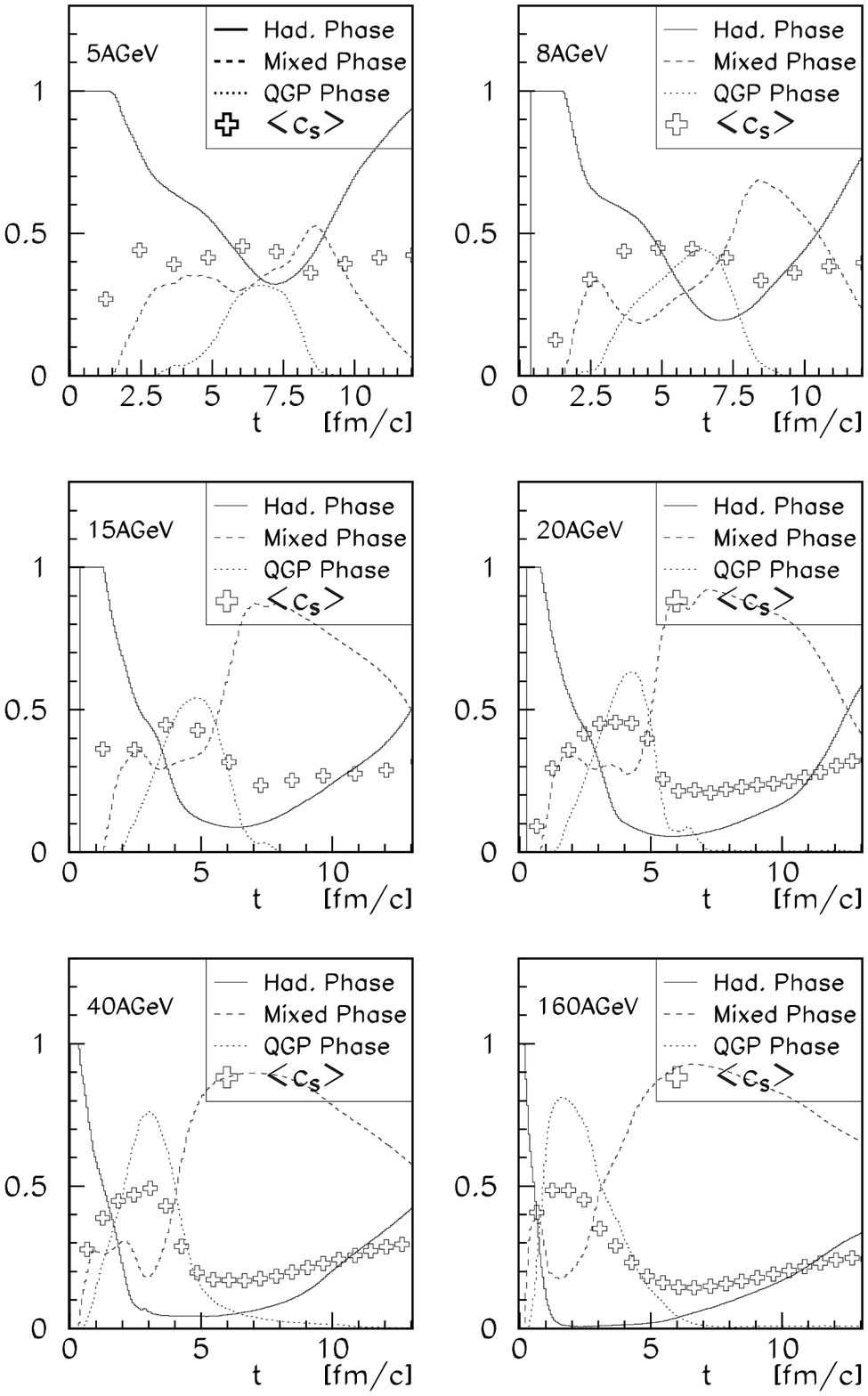,width=12cm}}}
\caption{Evolution of the isentropic speed of sound and the
fraction of baryon charge in the various phases;
Pb+Pb collisions at ${b=3}$~fm.}
\label{fig:plasma}
\vspace*{.5cm}
\end{figure}

Fig.~\ref{fig:plasma} shows the average isentropic velocity of sound, where
\be
c_s^2=\frac{\partial p}{\partial e}\Bigr|_{s/\rho}\quad,
\ee
and the fraction of baryon charge in each phase at various energies.
Throughout the manuscript, averages on fixed CM-time hypersurfaces employ
the time-like component of the net-baryon four-current
as the weight-function.

\begin{figure}[p]
\vspace*{-1cm}
\hspace{0cm}\centerline{\hbox
{\psfig{figure=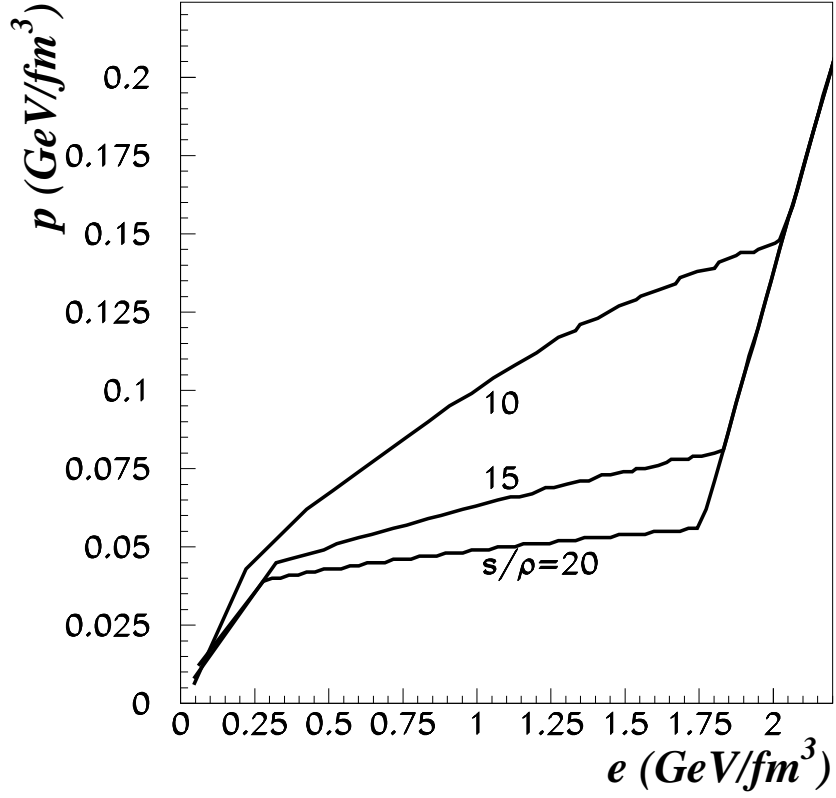,width=12cm}}}
\caption{Pressure as a function of energy density for various values
of the entropy per net baryon. $s/\rho=10$, 20 corresponds to
$E^{kin}_{Lab}=10A$~GeV and $40A$~GeV, respectively.}
\label{eps_n}
\vspace*{.5cm}
\end{figure}

The formation of plasma droplets starts already at BNL-AGS
energies. However, the baryon density $\rho$
in this energy domain is large and $\langle c_s\rangle$ is not very small.
Fig.~\ref{eps_n} shows the pressure as function
of energy density at fixed specific entropy. The values $s/\rho=10$, 20
correspond to $E^{kin}_{Lab}=10A$~GeV and $40A$~GeV \cite{SA_NPA}.
The derivative ($=c_s^2$) is only small within the mixed phase if the
specific entropy is large (hot matter).
Thus, the response of the highly excited matter to isentropic density
gradients is {\em not} weak at BNL-AGS energies.
However, at the higher CERN-SPS energies, $E^{kin}_{Lab}\ge40A$~GeV,
an extended time-interval where $\langle c_s\rangle$ is small {\em does}
exist.

From the above it is also clear that the previously
predicted~\cite{Yaris,Hung,Rischke:1996em} slow rehadronization of mixed phase
can only occur at rather high specific entropy, and thus bombarding
energy, but probably not at BNL-AGS energies. Indeed,
Fig.\ \ref{fig:plasma} shows that the maximum fraction of mixed phase occurs
always around a CM-time of $t\simeq7$~fm, despite the different energy scales.
Note also that non-equilibrium effects in the early stage of the
reaction limit the gain of QGP; it does not increase dramatically from
$E^{kin}_{Lab}=40A$~GeV to $160A$~GeV.

The EoS $p=p(e,\rho)$ is usually investigated by studying various flow 
patterns~\cite{LBCs,Yaris,Hung,xc60,Daniel,LPX,pxytheo,CR,afl3f}.
We shall discuss here how the directed in-plane flow, formerly called the
{\em bounce-off}~\cite{oldflow,Kampert}, ``responds'' to the
behavior of $\langle c_s\rangle$.

\begin{figure}[bp]
\vspace*{-1cm}
\hspace{0cm}\centerline{\hbox
{\psfig{figure=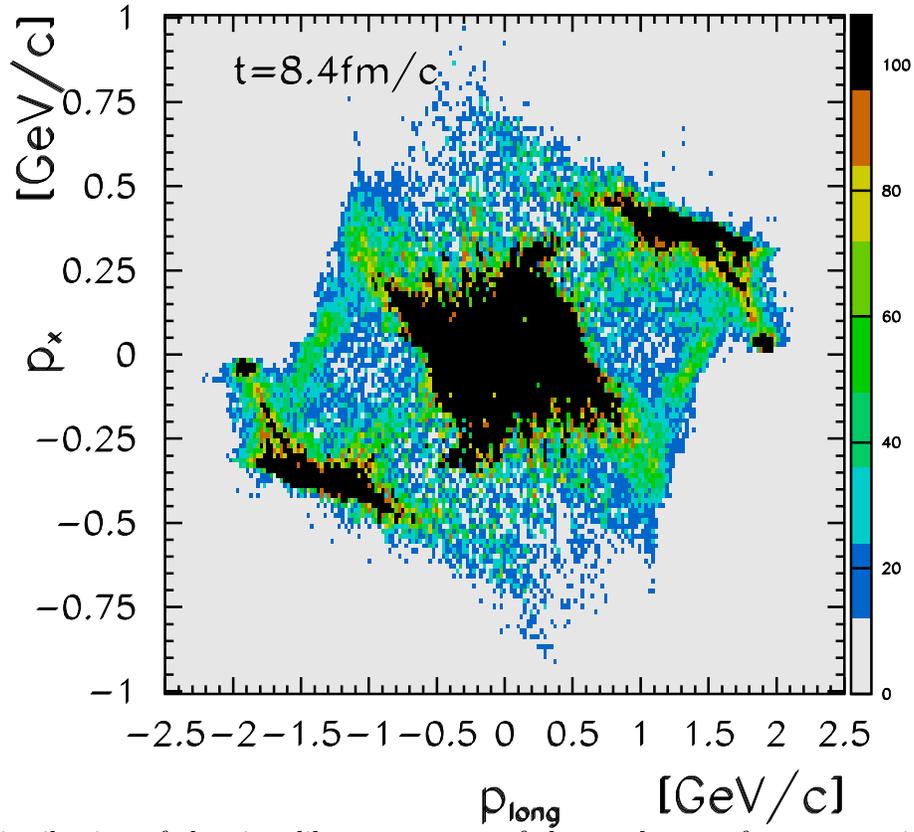,width=12cm}}}
\caption{Distribution of the time-like component of the net-baryon
four-current in momentum space ($p_x-p_{long}$ plane). The highest
contour-level is at 1/30th of the maximum, to offer a clearer view
of the midrapidity region.
Pb+Pb collisions at ${b=3}$~fm, $E^{kin}_{Lab}=8A$~GeV.}
\label{fig:pxy8gev}
\vspace*{.5cm}
\end{figure}

Fig.~\ref{fig:pxy8gev} shows the flow pattern obtained at
$E^{kin}_{Lab}\simeq8A$~GeV, at a time where we expect
departure from ideal flow in the central region.
We observe almost no net $p_x$
around midrapidity ($\left| p_{long}\right| <1$~GeV, say).
This is due to the expansion into the direction orthogonal to the
normal directed flow, resulting from the geometrical shape of the hot
and dense region in coordinate space~\cite{CR,afl3f}. Obviously, the
baryon number can only evolve towards larger isotropy in momentum space
because $\langle c_s\rangle$ is {\em not} small at this energy.

\begin{figure}[bp]
\vspace*{-1cm}
\hspace{0cm}\centerline{\hbox
{\psfig{figure=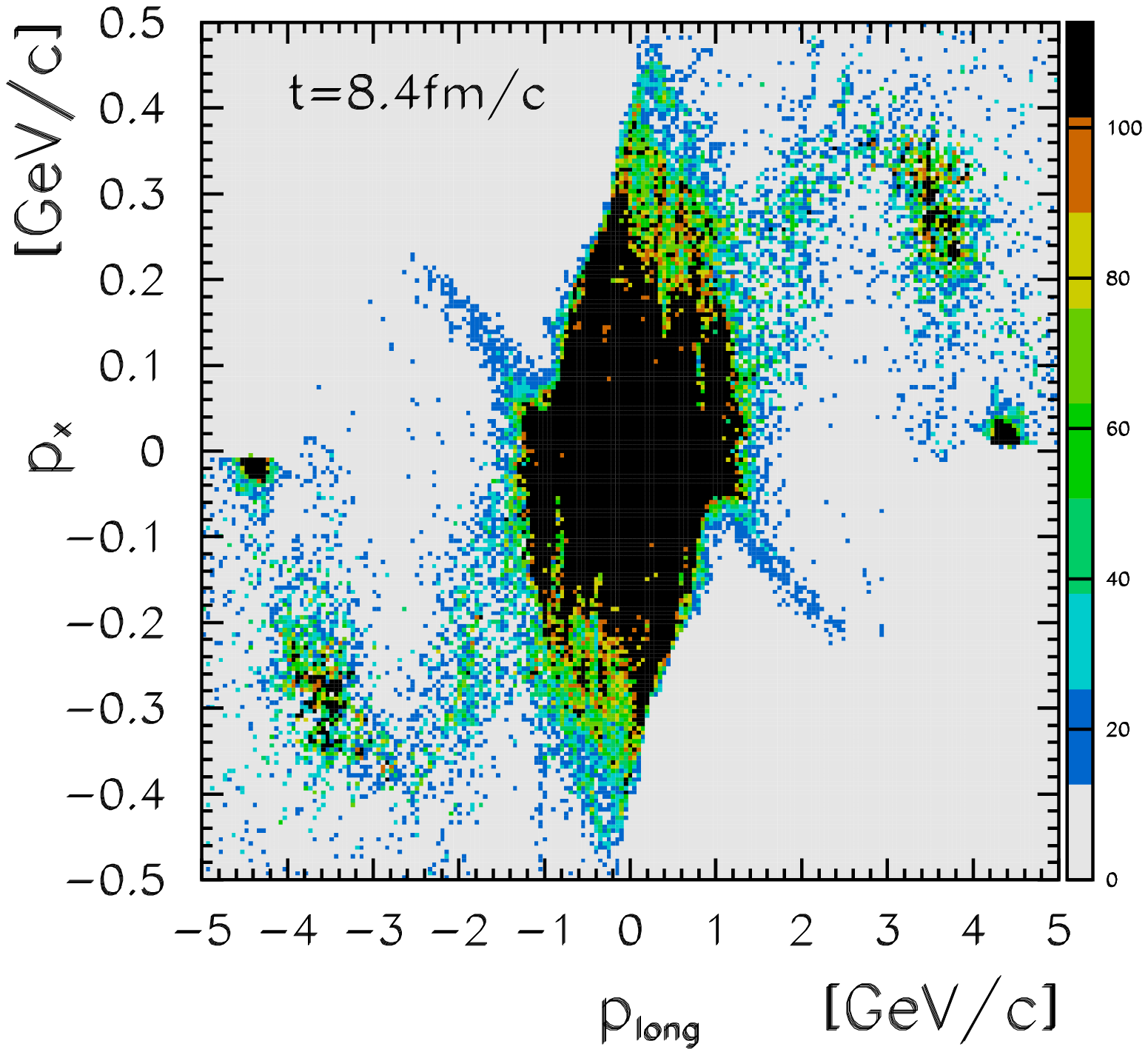,width=12cm}}}
\caption{As in Fig.~\ref{fig:pxy8gev} but for $E^{kin}_{Lab}=40A$~GeV.}
\label{fig:pxy40gev}
\end{figure}

At $E^{kin}_{Lab}=40A$~GeV, $\langle c_s\rangle$ is less than 0.2
when the phase coexistence region is reached from above.
This {\em prevents} more isotropic redistribution of the baryon number in
momentum-space, cf.\
Fig.~\ref{fig:pxy40gev}. The distribution is clearly very different from
that shown in Fig.~\ref{fig:pxy8gev}, with almost no
baryons in the upper left or bottom right quadrants where
$p_x\cdot p_{long}<0$.

At even higher energies, $E^{kin}_{Lab}=160A$~GeV and more,
the average $p_x$ per baryon becomes small again (around midrapidity)
for purely kinematical reasons~\cite{Yaris,LPX,afl3f}, as
can also be observed in the data~\cite{Harry}. That behavior is rather
insensitive to the EoS.

In summary, we investigated the excitation function of QGP formation
in relativistic heavy-ion collisions. Within our model
we find that droplets of QGP coexisting with hadronic
matter are already produced at the top BNL-AGS energy, 
$E^{kin}_{Lab}\simeq10A$~GeV. However, the average speed of sound in the
dense baryonic matter does not drop very much,
even in case of a first-order phase
transition. As a consequence, expansion of matter is not inhibited and
we find an almost isotropic momentum distribution in the reaction plane.
This is consistent with experimental results obtained for Au+Au
collisions at $E^{kin}_{Lab}=2-10A$~GeV \cite{AGSdata}, showing
that the net in-plane momentum diminishes as beam energy increases towards
top BNL-AGS energy.

Furthermore, we showed that the forthcoming results of the Pb+Pb
reactions at $E^{kin}_{Lab}=40A$~GeV will test whether the picture
of hot QCD-matter as a heat-bath with small isentropic speed of sound
(mixed phase) is indeed applicable to heavy-ion collisions
in this energy domain. {\em If} it holds true, our model calculation
predicts an increase of the directed net in-plane flow
around midrapidity as compared to top BNL-AGS energy, and a clearly
nonisotropic momentum distribution around midrapidity. Due to the
phase transition, pressure gradients along isentropes are too small to 
work towards a more isotropic momentum distribution.
Thus, those reactions can shed light on the (non-?)existence
of a local maximum of directed flow, related to a substantial decrease of
the isentropic speed of sound.

\acknowledgements
This work was supported by DFG, BMBF, GSI.
We thank L.P.\ Csernai, M.\ Gyulassy, I.N.\ Mishustin, D.H.\ Rischke,
and L.\ Satarov for numerous interesting discussions.
A.D.\ acknowledges support from the DOE Research Grant, Contract No.
De-FG-02-93ER-40764.

\newpage

\end{document}